\documentclass[prb,preprint]{revtex4}
\usepackage{graphicx,graphics,amssymb,amsmath}
\begin{document}
\title{Pinwheel VBS state and triplet excitations in the two-dimensional deformed kagome lattice}
\author{K. Matan$^{1,\ast}$, T. Ono$^2$, Y. Fukumoto$^3$ , T.~J.~Sato$^{1}$,  J. Yamaura$^4$, M. Yano$^2$, K. Morita$^2$, \& H. Tanaka$^2$}
\affiliation{
 $^1$Neutron Science Laboratory, Institute for Solid State Physics, University of Tokyo, 106-1 Shirakata, Tokai, Ibaraki 319-1106, Japan\\
 $^2$Department of Physics, Tokyo Institute of Technology, Meguro-ku, Tokyo 152-8551, Japan\\
 $^3$Department of Physics, Faculty of Science and Technology, Tokyo University of Science, Noda, Chiba 278-8510, Japan\\
 $^4$Institute for Solid State Physics, University of Tokyo, 5-1-5 Kashiwanoha, Kashiwa, Chiba 277-8581, Japan
}
\begin{abstract}
Determining ground states of correlated electron systems is fundamental to understanding novel phenomena in condensed matter physics.  A difficulty, however, arises in a geometrically frustrated system in which the incompatibility between the global topology of an underlying lattice and local spin interactions gives rise to macroscopically degenerate ground states\cite{ramirez}, potentially prompting the emergence of quantum spin states, such as resonating valence bond (RVB) and valence bond solid (VBS).  Although theoretically proposed to exist in a kagome lattice -- one of the most highly frustrated lattices in two dimensions (2D) being comprised of corner-sharing triangles -- such quantum-fluctuation-induced states have not been observed experimentally. Here we report the first realization of the ``pinwheel'' VBS ground state in the $S=\frac{1}{2}$ deformed kagome lattice antiferromagnet Rb$_2$Cu$_3$SnF$_{12}$.  In this system, a lattice distortion breaks the translational symmetry of the ideal kagome lattice and stabilizes the VBS state.
\end{abstract}
\maketitle

Theoretically, significant progress has been made toward understanding the ground state of the quantum $(S=\frac{1}{2})$ kagome lattice antiferromagnet via various approaches.  Diverse classes of novel states have been proposed; those include a gapless U(1)-Dirac-spin-liquid state\cite{Hastings:2001p247,Ran:2007p28}, a gapped-spin-liquid state\cite{Sachdev:1992p773,Waldtmann:1998p817,ZENG:1995p1054,Misguich:2002p3766}, and a VBS state\cite{MARSTON:1991p327,Nikolic:2003p879,Singh:2007p3664,Yang:2008p462}.  One leading approach is built upon a quantum dimer model, which was first introduced to describe spin dynamics of singlet pairs (dimers) in high-$T_c$ superconductors\cite{ROKHSAR:1988p4013,Moessner:2002p5988}.  Based on this model, two kinds of quantum states promptly emerge, namely the VBS and RVB states\cite{ZENG:1995p1054,Misguich:2002p3766,Nikolic:2003p879,Singh:2007p3664,Yang:2008p462}. The VBS state has long-range dimer-dimer correlations that break the translational symmetry of the ideal kagome lattice with spin-1 singlet-to-triplet excitations, whereas the gapped RVB state, in which different dimer coverings resonate among themselves, has short-range dimer-dimer correlations and topological order with continuum spin-$\frac{1}{2}$ excitations (spinons). Since these quantum states are non-magnetic and lack static order, one needs to look for their distinct features in magnetic excitations by means of inelastic neutron scattering to distinguish them in a real system.

Experimentally, realizations of these states have been extensively studied in quasi-1D\cite{FUJITA:1995p228,Nikuni:2000p3564,Lake:2005p1032} and Shastry-Sutherland 2D systems\cite{Kageyama:1999p2701}.  In contrast, the rarity of model systems has hitherto precluded experimental attempts to observe these states in the kagome lattice. In recent years, the discoveries of the kagome lattice in herbertsmithite\cite{Helton:2007p2715,Mendels:2007p335,Lee:2007p1219} and distorted kagome lattice in volborthite\cite{Bert:2005p1111,yoshida:077207} have been generating a lot of excitement and debate over their plausible ground states.  However, the lack of single crystals and Zn-Cu intersite disorder in herbertsmithite hinder the study of these systems. In this work, we present a single-crystal study of a stoichiometrically pure $S=\frac{1}{2}$ deformed kagome lattice antiferromagnet Rb$_2$Cu$_3$SnF$_{12}$ with full occupancy of spin-$\frac{1}{2}$ at the Cu sites.  Combined neutron scattering measurements on a large single crystal and advanced numerical analysis provide the first direct evidence for the pinwheel VBS state.

The magnetic $S=\frac{1}{2}$ Cu$^{2+}$ ions of Rb$_2$Cu$_3$SnF$_{12}$ reside at the corners of the triangles of the deformed single-layer kagome lattice, and are inside distorted octahedral cages formed by F$^{-}$ ions. At a room temperature, the system crystallizes in the hexagonal $R\bar{3}$ space-group with the lattice constants $a=13.917(2)$~\AA~ and $c=20.356(3)$~\AA\cite{Morita:2008p2717} (Figure~1{\bf a}). The slightly distorted kagome lattice yields four antiferromagnetic exchange interactions (Figure~1{\bf b}), which are characterized by different Cu$^{2+}-$F$^{-}-$Cu$^{2+}$ bond angles ranging from $124^\circ$ for the weakest bond to 138$^\circ$ for the strongest bond. The system does not exhibit magnetic order down to 1.3 K\cite{Morita:2008p2717,ono:174407}, which is confirmed by our elastic neutron scattering on a powder sample.  At low temperatures, magnetic susceptibility rapidly decreases toward zero, a signature of a non-magnetic, spin singlet $(S=0)$ ground state.  Furthermore, low temperature magnetization sharply rises when the applied field $H$ exceeds a critical field $H_c\approx20$ T, indicative of singlet-pair breaking and closing of a singlet-to-triplet gap. Finite magnetic susceptibility observed at the base temperature when the field is applied perpendicular to  the crystallographic $c$-axis is attributed to the mixing of the singlet and triplet states via the antisymmetric Dzyaloshinskii-Moriya (DM) interaction\cite{Morita:2008p2717}.  Therefore, to a first approximation, the spin Hamiltonian is given by:
\begin{equation}
{\cal H} = \sum_{nn} \left[ J_{ij} \textbf{S}_i \cdot \textbf{S}_j + \textbf{D}_{ij}\cdot \textbf{S}_i \times \textbf{S}_j \right],\label{model_H}
\end{equation}
where $J_{ij}$ are the nearest-neighbor interactions for four different bonds $J_1>J_2>J_3>J_4$ as labelled in Figure~1{\bf b}, and \textbf{\textit{D}}$_{ij}=\textbf{\textit{D}}_{1}$, \textbf{\textit{D}}$_{2}$, \textbf{\textit{D}}$_{3}$, and \textbf{\textit{D}}$_{4}$ are the DM vectors, whose in-plane components point toward the center of a triangle and out-of-plane components alternately point into and out of the kagome plane\cite{ono:174407}. The spatial distribution of the non-uniform exchange interactions suggests the stabilization of the pinwheel VBS state\cite{Syromyatnikov:2002p244} in which singlet dimers are formed between two spins bonded by $J_1\sim20$ meV\cite{Morita:2008p2717}.  This pinwheel VBS state is different from the VBS state theoretically proposed for the ideal kagome lattice, since it is stabilized by the lattice distortion and  the long-range dimer-dimer correlations do not break the translational symmetry of the lattice.  A theoretical study of the $J_4$-depleted kagome lattice\cite{yang:224417}, which is defined by $J_1=J_2=J_3\equiv J$ and $J_4\equiv\alpha J$,  shows that the most energetically-favored VBS state proposed for the ideal kagome lattice, which corresponds to $\alpha=1$, is destabilized against this pinwheel state even at $\alpha\sim0.97$, highlighting the importance of geometrical frustration which leads to the diversity and closeness of exotic phases in the quantum kagome lattice. 

To verify the pinwheel VBS state, we performed inelastic neutron scattering on a single crystal sample (see Methods). The scattering intensity can be described by the following equation:
%\begin{widetext}
\begin{equation}
I({\bf Q},\hbar\omega)=\left(\frac{\gamma r_0}{2}\right)^2f({\bf Q})^2S(\hbar\omega,{\bf q; {\bf Q}_m}),\label{eqInt}
\end{equation}
%\end{widetext}
where $\left(\frac{\gamma r_0}{2}\right)^2=72.65\times10^{-3}$ barn$/\mu_B^2$, $f({\bf Q})$ is a magnetic form factor for Cu$^{2+}$, $S(\hbar\omega,{\bf q}; {\bf Q}_m)$ is a dynamical structure factor, and ${\bf Q}={\bf Q}_m+{\bf q}$. An energy scan at the Brillouin zone center ${\bf Q}_m=(0,2,0)$ (Figure~2{\bf d}) clearly shows two singlet-to-triplet gaps. The convoluted fit with the resolution function yields the gap energies $\Delta_1=2.35(7)$ meV and $\Delta_2=7.3(3)$ meV.  The thermal renormalization of the spectral weight and damping of the spin excitations (Figure~3) are typical of triplet modes in an interacting dimer system\cite{Ruegg:2005p1755}.  In addition, energy scans in an applied magnetic field (Figure~4) reveal the Zeeman splitting of the lower gap and no splitting of the upper gap. These results demonstrate that, whereas the upper gap is a result of the excitations from the singlet to $S_\textrm{tot}^z=0$ triplet states, the lower gap, which is two-fold degenerate at zero field, is due to the excitations from the singlet to $S_\textrm{tot}^z=\pm1$ triplet states $(S^z_{\textrm{tot}}$ represents a triplet spin state of a single dimer). The Zeeman splitting of the $S_\textrm{tot}^z=\pm1$ triplet gap can be described by the following relation $\Delta_{1,\pm}(H)=\Delta_1- g_\pm\mu_BS^z_{\textrm{tot}}H$, where $\mu_B$ is the Bohr magneton. For $S_\textrm{tot}^z=1$ and $-1$, we obtained $g_+=2.58(30)$ and $g_-=2.34(20)$, respectively. Extrapolating the $S_\textrm{tot}^z=1$ line to intercept the $H$-axis, i.e. $\Delta_{1,+}(H=H_c)=0$, yields the critical field $H_c\approx 21$~T, consistent with the magnetization data\cite{Morita:2008p2717}.  The magnitude of the DM vector can be estimated from Moriya's calculation\cite{Moriya:1960p1210} as $|\vec{D}|/J_{av}\sim\Delta g/g\sim 0.2$ ($J_{av}\equiv\frac{1}{4}\sum_{i}J_i$).  Both Land\'e $g$-factor and $|\vec{D}|$ are in good agreement with the values obtained from fitting the susceptibility data to the exact diagonalization for the 12-site kagome cluster\cite{Morita:2008p2717,ono:174407}.

For Rb$_2$Cu$_3$SnF$_{12}$, six dimers in the unit cell give rise to  six triplet modes.  In the limit $J_2-J_4\rightarrow0$, these modes can be characterized by irreducible representations of the cyclic group $C_3$ as four E and two A localized (flat) modes (see Figure 6 in ref. 26 and Figure S3 in the Supplemental Information).  The observed triplet excitations belong to the lowest-energy E mode. The partial lifting of this E-mode triplet degeneracy can be attributed to the DM interaction. For a single dimer, one can easily show that the DM interaction raises the energy of the $S_\textrm{tot}^z=0$ state, while retaining the degeneracy of the two $S_\textrm{tot}^z=\pm1$ states. To explain the observed large bandwidth and small $S_\textrm{tot}^z=\pm1$ triplet gap, we performed the dimer series expansion on the pinwheel VBS ground state using the model Hamiltonian given in equation (1). We assume that the strength of the DM vectors scales with the respective exchange interactions, \textit{i.e.} $d_z=D^z_i/J_i$ ($d_p=D^p_i/J_i$) where $i=$1, 2, 3, and 4 for the out-of-plane (in-plane) component. In our first-order dimer expansion (see Figure S3), the out-of-plane component $d_z$ of the DM interaction lowers the $S_\textrm{tot}^z=\pm1$ branch at $\Gamma$ point, which strongly suggests that the observed two branches are the lowest-energy E-mode in the $S^z_{tot}=\pm1$ and 0 sectors.  On the other hand, the in-plane component $d_p$ only determines detailed structure of further energy splitting, and thus is omitted in the higher-order calculations.  The linked cluster expansion algorithm\cite{oitmaa} was used to obtain a longer series.  The lowest-energy triplet excitation spectra were then calculated using the Dlog-Pad\'e approximation up to eighth order in the inter-dimer exchange and DM interactions.  Under the constraints $J_1>J_2>J_3>J_4$ and a fixed $J_{av}$, we obtain the best fit to the excitation spectra (Figures~2{\bf a}-{\bf c}) for $J_1=18.6$ meV, $J_2=0.95J_1$, $J_3=0.85J_1$, $J_4=0.55J_1$, and $d_z=0.18$. The agreement between the measured and calculated triplet dispersions verifies the pinwheel VBS state. The integrated dynamic structure factor (Figure 2{\bf e}) was calculated to fifth order, assuming equal  contributions from two domains, one domain with pinwheel arms in clockwise direction and the other with pinwheel arms in the counterclockwise direction. These two pinwheel motifs correspond to two equivalent structural domains, which are a mirror image of each other (see Figure S5 in Supplemental Information).  The intensity pattern with the periodicity four times the reciprocal lattice vector correlates with the intra-dimer distance, which is one forth of the in-plane lattice constant. The maximum scattering intensity around (2,~0,~0), (0,~2,~0) and (2,~2,~0) also supports the pinwheel VBS state in Rb$_2$Cu$_3$SnF$_{12}$ (see Supplemental Information).

In contrast to triplet excitations in other dimer systems, the highly-dispersive triplet bands and the small triplet gap ($\Delta_1/J_1\sim0.13$) in Rb$_2$Cu$_3$SnF$_{12}$ are due to the strong inter-dimer interactions, the DM interaction, and the unique pinwheel arrangement of dimers. The bandwidth of the $S_\textrm{tot}^z=0$ branch is denominated by $J_2-J_4$; therefore, the rather small $J_4\approx J_2/2$ is essential to explain its relatively large bandwidth.  It is also worth noting that the $S_\textrm{tot}^z=0$ branch is virtually unaffected by $d_z$ and has a gap minimum at {\bf K} point, as predicted by the bond operator mean field theory without the DM interaction\cite{yang:224417}. On the other hand, $d_z$ changes the gap minimum of the $S_\textrm{tot}^z=\pm1$ branch from {\bf K} point to $\Gamma$ point, and significantly lowers the gap energy.  If the strength of $d_z$ exceeded $0.25$, the gap at $\Gamma$ point would be closed and magnetic order would instead be observed. 

The macroscopically degenerate ground states in the kagome lattice are particularly sensitive to small perturbations, and thereby could inherently give rise to a number of different states in a real system. In Rb$_2$Cu$_3$SnF$_{12}$, our results reveal the pinwheel VBS state that has been suggested to be in close proximity to other phases\cite{Yang:2008p462,yang:224417}. Recently, we have studied other related compounds, in which a different ground state was observed\cite{ono:174407}.  In particular, Cs$_2$Cu$_3$SnF$_{12}$ exhibits magnetic order but possesses the low-temperature phase that shares the crystal structure with Rb$_2$Cu$_3$SnF$_{12}$, which suggests similar magnetic interactions. The magnetic order in Cs$_2$Cu$_3$SnF$_{12}$ could, therefore, be ascribed to (i) a small difference in exchange coupling and/or (ii) a large DM interaction\cite{Cepas:2008p5529}.  In this case, the former could play a more prominent role in determining the ground state since the DM interactions in these systems are essentially comparable\cite{ono:174407}. In addition, other phases could be obtained by tuning external parameters such as magnetic field and pressure.  Further experimental and theoretical studies are desirable to determine precisely the driving mechanism that leads to these different ground states.  A more comprehensive understanding of this mechanism could shed some light on many remaining questions regarding the quantum kagome lattice, particularly in relation to the sought-after quantum spin liquid.

\vspace{5 mm}
{\bf Methods}
\vspace{5 mm}

{\bf Experiments}
\vspace{5 mm}

{\small Inelastic neutron scattering experiments were preformed on the HER and GPTAS triple-axis spectrometers operated by the Institute for Solid State Physics, University of Tokyo.  Vertically focusing pyrolytic graphite (PG) crystals were used to monochromate the incident neutron beam.  For the measurements on the HER spectrometer, the scattered neutrons with a fixed final energy of 5 meV were analyzed by the central three blades of a seven-blade doubly focused PG analyzer.  A cooled Be or oriented-PG-crystal filter was placed in the incident beam, and a room-temperature Be filter in the scattered beam to remove higher-order contamination.  For the measurements on the GPTAS spectrometer, vertically focused (horizontally flat) PG crystals were used to analyze scattered neutrons with fixed final energies of 14.7 meV and 13.7 meV.  The horizontal collimation sequence of $40'-80'-$sample$-80'-80'$ (LR) or $40'-40'-$sample$-40'-80'$ (HR) was employed with a PG filter placed in the scattered beam.  A single crystal of mass 0.94~g was mounted in the $hk0$ and $h0l$ zones. The single crystal sample used in this study was synthesized using the method described in ref.~23. The sample was sealed in an aluminum container in a He gas environment for heat exchange, and cooled by a closed cycle $^4$He cryostat. For the measurements in a magnetic field, the sample was aligned in the $hk0$ zone, and the field was applied along the crystallographic $c$-axis, perpendicular to the kagome plane, using a vertical magnet with a maximum field of 5 T. The sample was then cooled by a liquid $^4$He cryostat equipped with the magnet.}

\vspace{10 mm}
{\bf Data Analysis}
\vspace{10 mm}

{\small In Figure~2{\bf a}, the measurements were done around (2,~0,~0) for energies between 2 meV and 4.5 meV, and around (2,~2,~0) between 4.5 meV and 9 meV. In Figure~2{\bf b}, the measurements were done around (2,~2,~0) along two equivalent high symmetry directions; for energies between 2 meV and 4 meV, the scans were taken along (H,~H,~0), and between 4 meV and 9 meV along (H,~6-2H,~0). In Figure~2{\bf a}, the triplet branch around (1.5,~0,~0) could be a result of higher-order corrections to the exchange interactions due to small atomic displacements as evidenced by weak superlattice reflections; we have ruled out the contribution of other triplet modes (see Supplemental Information). All scans were analyzed using an empirical dispersion relation, $\hbar\omega_{\alpha}(\textbf{q})=\sqrt{\Delta_{\alpha}^2-(-1)^\alpha (v_{\alpha,x}^2k_x^2+v_{\alpha,y}^2k_y^2)}$, where $\alpha =1$ or $2$ for two non-degenerate branches, and ${\bf q}=(k_x,k_y)$ is a wave vector away from a Brillouin zone center ${\bf Q}_m$.  Within the resolution of the instrument of $\sim1$ meV (GPTAS-LR), the out-of-plane dispersion is flat, attesting to the two-dimensionality of the system, and hence we set $v_{\alpha,z}=0$. The observed peaks were fit with narrow, resolution-limited Lorenztians, convoluted with the resolution function taking into account the empirical dispersion relation. The line shapes are governed by the convolution with the resolution function.  The peak positions in a unit  of~\AA$^{-1}$ were obtained by fitting $v_\alpha$ and an offset $\Delta {\bf q}$ (see Figure~S2 in Supplemental Information).}

\vspace{10 mm}
We thank M. Nishi, C. Broholm, K. Iida, and M. Kofu for useful discussions. The work at Tokyo Institute of Technology was supported by a Grant-in-Aid for Scientific Research from JSPS and a Global COE Program funded by MEXT Japan.

%\bibliographystyle{naturemag}
%\bibliography{RCSF-nature}

\vspace{10 mm}
{\small {\bf Author Contributions:} K. Matan, T. O., and T. J. S. performed neutron scattering measurements.  J. Y. performed X-ray diffraction measurements. Y. F. preformed numerical calculations. T. O., M. Y., K. Morita, and H. T. contributed to sample synthesis and characterization.}

{\small {\bf Competing Interests:} The authors declare that they have no competing financial interests.

{\small {\bf Correspondence:} Correspondence and requests for materials should be addressed to K. Matan~(kmatan@issp.u-tokyo.ac.jp), T. O. (ono.t.aa@m.titech.ac.jp), and Y. F. (yfuku@ph.noda.tus.ac.jp).

{\small $^\ast$Present address: Department of Physics, Faculty of Science, Mahidol University, 272 Rama VI Rd., Ratchathewi, Bangkok 10400 Thailand.
\newpage

\begin{figure}
\centering \vspace{0in}
\includegraphics[width=15cm]{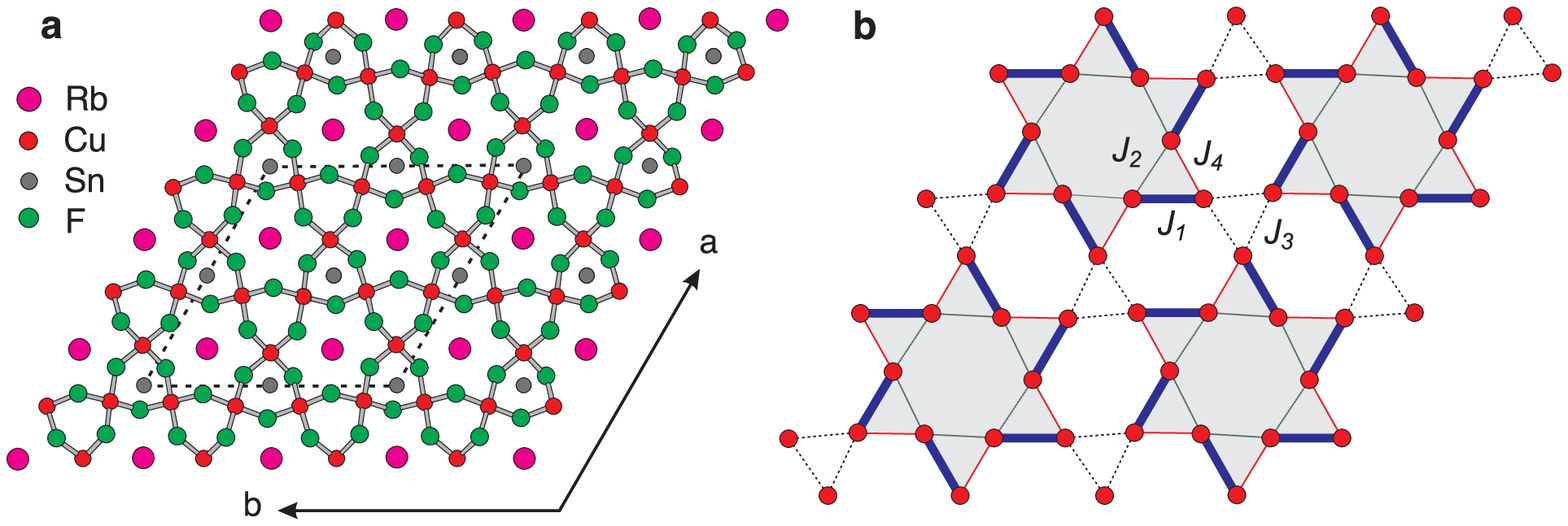}
\caption{Crystal structure of Rb$_2$Cu$_3$SnF$_{12}$. {\bf a} A crystal structure in the $ab$ plane shows the connectivity of the Cu$^{2+}$ ions (red) forming a deformed kagome lattice.  In {\bf b}, dimers (blue bonds) form the pinwheel VBS state.  The exchange interactions are labelled as $J_1>J_2>J_3>J_4$.}\label{fig1}
\end{figure}
\clearpage

\begin{figure}
\centering \vspace{0in}
\includegraphics[width=15cm]{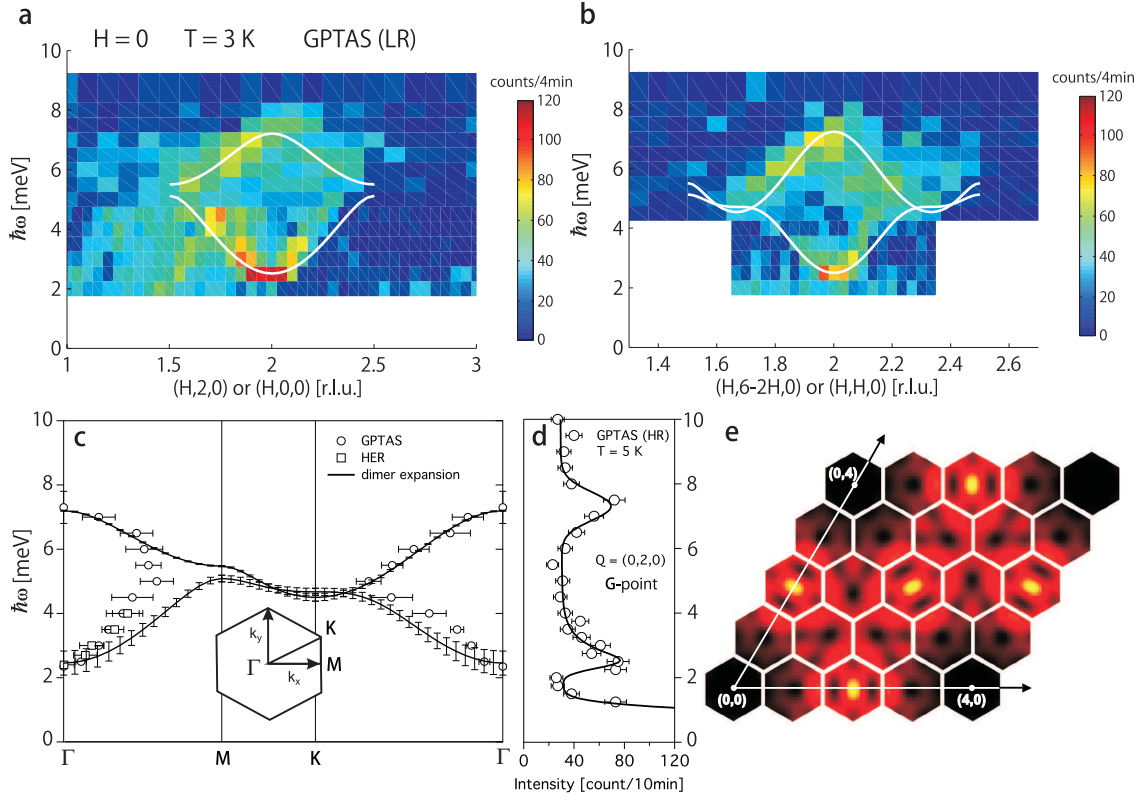}
\caption{Energy-momentum contour maps.  Intensity contour maps were measured at $T = 3$ K around (2, 0, 0) and (2, 2, 0) along two high symmetry directions, {\bf a} along $k_x$ and {\bf b} along $k_y$.   {\bf c} Triplet dispersions were measured along the path shown in the inset.  Solid lines represent the best fits to the dimer series expansion described in the text. Error bars reflect the differences of various Pad\'e approximations.  {\bf d} An energy scan at ${\bf Q}_m =$ (0, 2, 0) ($\Gamma-$point) and $T= 5$ K shows two spin gaps at 2.35(7) meV and 7.3(3) meV. {\bf e} The calculated dynamic structure factor integrated over triplet excitations shows high scattering intensity around (2, 0, 0), (0, 2, 0) and (2, 2, 0), consistent with the experimental data.}  \label{fig2}
\end{figure}
\clearpage

\begin{figure}
\centering \vspace{0in}
\includegraphics[width=15cm]{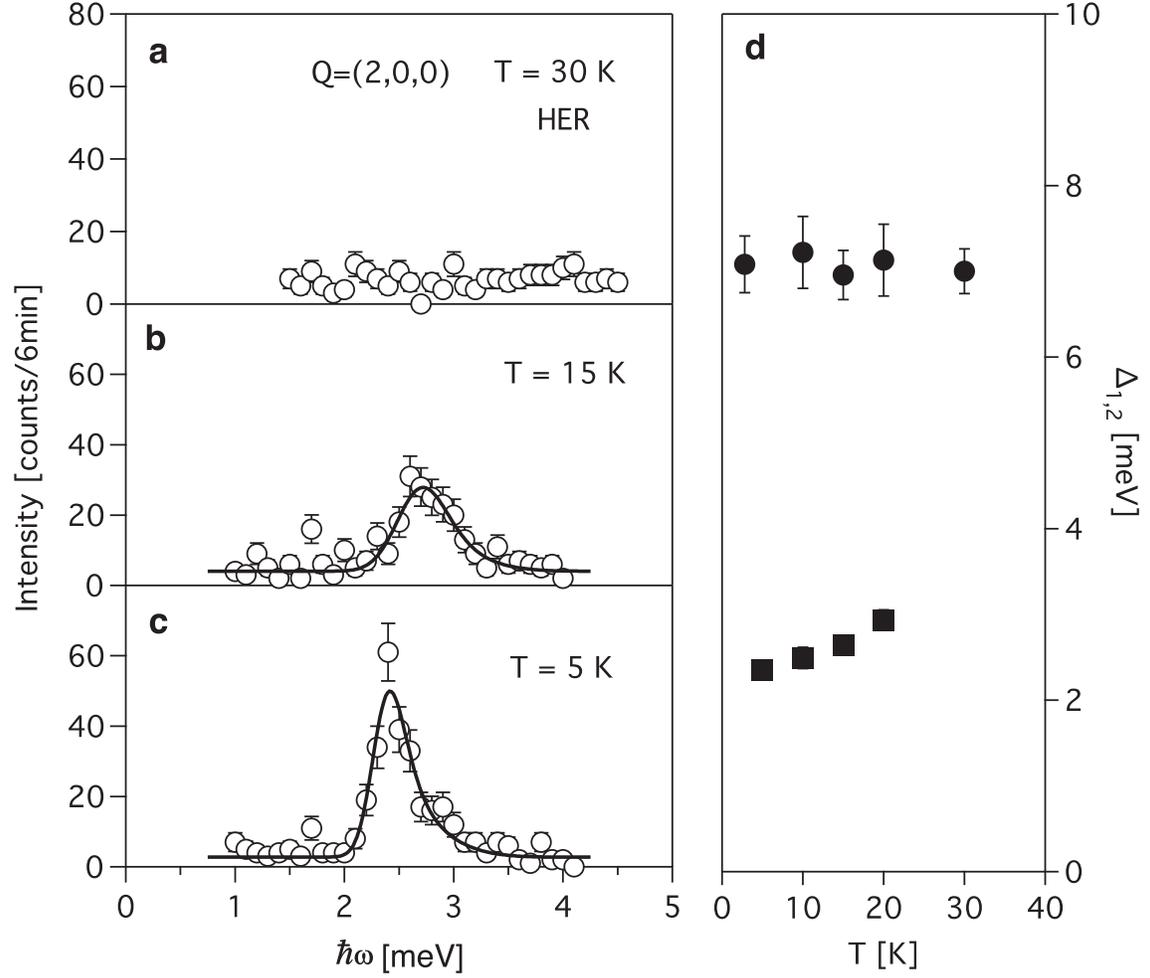}
\caption{Representative energy scans and temperature dependence.  {\bf a}-{\bf c} Energy scans of the lower gap at three temperatures. Solid lines denote the fits to the dispersion relation described in Methods, convoluted with the instrumental resolution function. Peak widths are resolution-limited at the base temperature and become broader with increasing temperatures. Temperature dependence of the two triplet gaps is shown in {\bf d}.}\label{fig3}
\end{figure}
\newpage

\begin{figure}
\centering \vspace{0in}
\includegraphics[width=15cm]{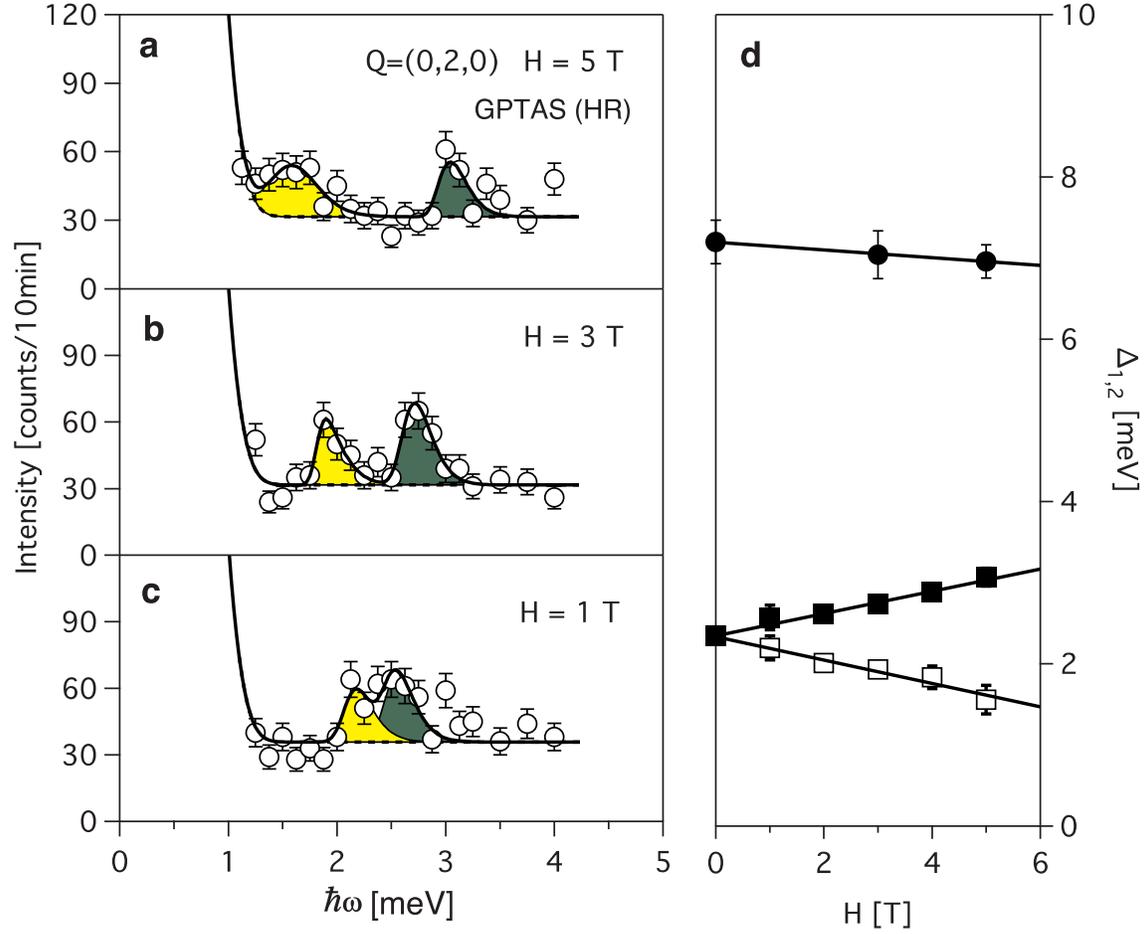}
\caption{Triplet excitations in a magnetic field. {\bf a}-{\bf c} Energy scans measured at $T = 5$ K show the splitting of the lower gap at three different fields. Solid lines denote the fits to the dispersion relation described in Methods, convoluted with the instrumental resolution function. Field dependence of the triplet gaps is shown in {\bf d}.}\label{fig4}
\end{figure}
\clearpage

\begin{center} {\large {\bf Supplemental Information}} \end{center}

\begin{center} {\bf Superlattice reflections} \end{center}

	Below $T_s \sim 215$ K, X-ray diffraction shows superlattice reflections at positions displaced from the fundamental Bragg peaks by commensurate wave vectors $(\pm1/2,\pm1/2,0)$, $(\pm1/2,\mp1,0)$, and $(\pm1,\mp1/2,0)$ (Figure S1(a)). The largest intensity of the superlattice peaks is roughly 50 times smaller than the fundamental Bragg peaks, indicative of small atomic displacements. The absence of an anomaly in magnetic susceptibility at $T_s$, as well as the relatively large superlattice intensity compared to that in other inorganic spin Peierls systems such as CuGeO$_3$ and TiOCl, suggests that this structural phase transition is not the spin Peierls transition. Hysteretic behavior of the superlattice intensity measured by neutron scattering (Figure S1(b)) suggests a first order structural transition at $T_s$.  Below $T_s$, the hexagonal lattice is retained but the $R\bar{3}$ space-group changes to $P\bar{3}$ or $P3$ with an enlarged unit cell. 

\begin{center} {\bf Triplet excitations around (1.5, 0, 0) and (2.5, 0, 0)} \end{center}

Dispersive branches of the triplet modes were also observed around (1.5, 0, 0) and (2.5, 0, 0) (see also Figure 2a).  A long constant-energy scan around (2, 0, 0) (Figure S1(c)) shows triplet excitations around (1.5, 0, 0) and (2.5, 0, 0), which disappear at high temperatures. Figure S1(d) shows the singlet-to-triplet gap at (1.5, 0, 0) with $\Delta_1 = 2.21(9)$ meV, consistent with the triplet gap at (2, 0, 0) within the experimental uncertainties.  The presence of these extra triplet branches and the spin gap at (1.5, 0, 0) indicates the change in the periodicity of the magnetic exchange interactions due to the enlarged unit cell of the low-temperature phase as evidenced by the superlattice reflections.  The small atomic displacements at low temperatures slightly alter the exchange-interaction network, and thereby give rise to a new periodicity of the triplet modes.  The scattering intensity of the triplet modes around (1.5, 0, 0) and (2.5, 0, 0) is about five times smaller than that around (2, 0, 0).  The fit in Figure S1(c) already takes in account the magnetic structure factor of the Cu$^{2+}$ ion. In order to account for this half periodicity, one needs to perform the dimer series expansion on 24 dimers or 4 pinwheels in the enlarged unit cell, which is beyond the scope of this study.  However, this correction is small, and does not change the pinwheel arrangement of dimers. The origin of this weak branch will be further discussed in the next section.

\newpage
\begin{center} {\bf Calculations of triplet excitation energies at M point} \end{center}

The possibility that the observed dispersive branches around the M points (1.5, 0, 0) and (2.5, 0, 0) are due to other triplet modes is ruled out as described below. 

We calculated the whole triplet excitation energies at M point using the dimer series expansion method up to eighth order. For E modes, we obtained reasonable convergence of the Dlog-Pad{\'e} approximations. The two lowest energies of the E modes with $S^z_{tot} = 0 (\pm1)$ at M point are 5.5 and 6.5 meV (5.1 and 7.3 meV), which are much higher than 2.4 meV. For A modes, a bare energy series indicates the tendency of softening at M point as a function of $\lambda$, which denotes a scaling parameter for the inter-dimer and DM interactions, i.e. $J_i = \alpha_i\lambda J_1$ where $i = 2, 3, 4$ and $\alpha_i$'s are constants. In order to check if it is a second order phase transition, we examined the Dlog-Pad{\'e} approximations of the eighth order series, in which we found that most of the Dlog-Pad{\'e} approximations have poles and the critical exponent for the vanishing gap is estimated as $\sim0.2$. Since this critical exponent is much smaller than the known gap exponent of 0.71 for two-dimensional systems, we conclude that the softening found in the eighth order series of the A mode energies is artificial due to the shortness of the series. We have no reliable energy estimations of the A modes because of the aforementioned artificial softening. However, our structure factor calculation indicates that scattering intensities of the A modes are much smaller than those of the E modes, which rules out the possibility that the scattering intensity at M point in the constant energy scan with $\hbar\omega\sim2.4$ meV comes from the A modes.  Therefore, we conclude that the extra mode at (1.5, 0, 0) and (2.5, 0, 0) is due to the magnetic super-structure, which corresponds to the enlarged structural unit cell evidenced by the superlattice reflections.

\begin{center} {\bf Representative constant-energy scans and dispersion relations (Figure S2)} \end{center}

Constant-energy scans were performed around (2, 2, 0) at T = 3 K with (a) $\hbar\omega = 3.5$ meV along $k_y$, (b) $\hbar\omega = 3.0$ meV along $k_x$, and (c) $\hbar\omega = 6.5$ meV along $k_x$. The solid lines show the fits to the dispersion relation described in Method, convoluted with the instrumental resolution function. (d) and (e) show dispersions of the singlet-to-triplet excitations along $k_x$ and $k_y$, respectively. In (d), the HER data are shown in green diamonds.  The dotted lines represent the model dispersion described in the text with a universal $v_{\alpha,i} = 35$ meV/\AA for $\alpha=1, 2$ and $i=x, y$.

\newpage
\begin{center} {\bf Characteristic features of the energy spectra} \end{center}

We here present the calculated energy spectra within the first-order perturbation, for comparison with the bond-boson calculation employed in ref. 26.  This lowest-order perturbation calculation provides comprehensive explanations for some characteristic features of the excitation spectra of the pinwheel VBS state.

We first focus on the exchange interaction, omitting the DM interaction, and describe how the dispersive modes emerge.  As pointed out in ref. 26, when $J_2=J_4$, a pinwheel, hatched regions in Figure 1b, becomes an ÔorthogonalÕ dimer ring, which prohibits a triplet dimer from propagating.  Hence, the spectra are non-dispersive (flat) as shown by black solid (E-mode) and dashed (A-mode) lines in Figure S3(a), since a triplet dimer is confined around a triangle formed by $J_3$.  The energies of E- and A-modes are given by $J_1-\frac{J_3}{4}$ and $J_1+\frac{J_3}{2}$, respectively.  A non-zero value of $J_2-J_4$ leads to the breakdown of this orthogonal nature and the triplet-excitation spectra become dispersive as shown by red curves in Figures S3(a).
	
We now turn to the effect of the DM interaction.  We first consider only the out-of-plane component $d_z$, where $S_{tot}^z$ is a conserved quantity.  Within the first-order perturbation, the effective Hamiltonians for the $S_{tot}^z=\pm1$ sectors depend on $d_z$ but that for the $S_{tot}^z=0$ sector does not.  The calculated spectra for $d_z=0.09$ are shown in Figure S3(b).  The $S_{tot}^z=±1$ branches (blue curves) are affected by $d_z\neq0$, whereas the $S_{tot}^z=0$ branches (red curves) are unchanged.  Figure S3(c) shows the result of $d_z=0.18$, in which the two lowest-energy branches denoted by thick lines resemble the measured dispersion curves.  We finally include the in-plane component $d_p$; the results for $d_z=d_p=0.18$ are shown by green lines in Figure S3(c).  We found that $d_p$ only determines detailed structure of the spectra and has a very small effect on the two lowest-energy branches. 

\begin{center} {\bf Dynamic structure factor} \end{center}

The dynamic structure factor $S(q,\omega)$ is defined by

\begin{equation}
S\left(\overrightarrow{q},\omega\right)=\frac{1}{2\pi N}\sum_{i,\xi}\sum_{i',\xi'}\int^\infty_\infty dt e^{i\omega t+i\overrightarrow{q}\cdot\left(\overrightarrow{r}_{i,\xi}-\overrightarrow{r}_{i',\xi'}\right)}\left<\Psi_g\left|\overrightarrow{S}_{i',\xi'}(t)\cdot\overrightarrow{S}_{i,\xi}\right|\Psi_g\right>
\end{equation}

where N denotes the total number of spins, $r_{i,\xi}$ the position of the $\xi^{th}$ spin on the $i^{th}$ unit cell, and $\psi_g$ the ground state.  Introducing a complete set of momentum eigenstates $\left|\psi(\overrightarrow{q})\right>$ of the Hamiltonian, we can rewrite the dynamic structure factor as $S(\overrightarrow{q},\omega)=\sum_\Lambda S_\Lambda(\overrightarrow{q},\omega)$ with $S_\Lambda(\overrightarrow{q},\omega)=I_\Lambda(\overrightarrow{q})\delta(\omega-E_\Lambda(\overrightarrow{q})+E_g)$.  In order to obtain the perturbative estimation of $S_\Lambda(\overrightarrow{q},\omega)$, we first utilize the linked cluster expansion algorithm to calculate series expansions of the exclusive matrix elements (see ref. 29) and the effective Hamiltonian $H_{eff}$. The energy $E_\Lambda(\overrightarrow{q})$ is obtained as an eigenvalue of $H_{eff}$, and the weight $I_\Lambda(\overrightarrow{q})$ is calculated from the exclusive matrix elements and an eigenvector of $H_{eff}$.

In Figure 2(e), we show the fifth order result of the integrated structure factor of triplet dimer (TD) states, $I_{TD} (\overrightarrow{q})=\sum_{\Lambda\in TD}\int d\omega S_\Lambda(\overrightarrow{q},\omega)=\sum_{\Lambda\in TD}I_\Lambda(\overrightarrow{q})$. Since we carry out numerical diagonalization of the fifth-order effective Hamiltonian to get eigenvectors required in the weight calculation, our calculated $I_\Lambda(\overrightarrow{q})$ is not in the form of a power series and thus we do not perform series analysis such as Pad{\'e} approximation.  Although our weight calculation contains no extrapolation scheme, it can reproduce the experimental feature of high scattering intensity in the Brillouin zones centered at (2, 0, 0), (0, 2, 0), (2, 2, 0) and so on, consistent with the experimental result.

Now we describe how the spectral weight $I_{TD}$ is allocated to individual triplet-dimer modes. Figures S4(a) and (b) show the calculated structure factor $S(q, \omega)$ along two equivalent high symmetry directions ($k_x$), (H, 0, 0) and (H, 2, 0), respectively. Similarly, Figures S4(c) and (d) show the calculated structure factor along the other two equivalent high symmetry directions ($k_y$), (H, H, 0) and (H, 6-2H, 0), respectively. These high symmetry directions correspond to the measuring paths shown in Figures 2a and b. We find four branches in the measuring energy range ($\hbar\omega < 9$ meV); we call them, from the bottom, first, second, third, and fourth branches. The first and fourth (second and third) branches belong to $S_z = \pm1(0)$ mode, and are denoted by dashed (dotted) lines.  In Figure 2a, the measurements were done for 2 meV $<  \hbar\omega <$ 4.5 meV along the (H, 0, 0) direction and for 4.5 meV $<  \hbar\omega <$ 9 meV along the (H, 2, 0) direction.  In Figure 2b, the measurements were done for 2 meV $<  \hbar\omega <$ 4 meV along the (H, H, 0) direction and for 4 meV $<  \hbar\omega <$ 9 meV along the (H, 6-2H, 0) direction.  These measuring ranges are indicated by white rectangles in Figure S4.  
	
In Figures S4(a) and (b), the presence of the intense first branch is in agreement with the experiment.  However, the weak second branch and intense third branch in (b) are inconsistent with the experiment.  Note that in the dispersion fitting we have assigned the observed higher-energy mode to the second branch.  Furthermore, while the second branch appears more intense than the third branch in (a), it becomes less intense in (b). (Similar inconsistencies were also observed along $k_y$ in Figures S4(c) and (d), though they are less obvious). These results imply that (2, 0, 0) and (2, 2, 0) are singular points of $S(q, \omega)$. This comes from the fact that the second and third branches are degenerate at $\Gamma$ point.  For $d_z$ = 0, all four branches converge at $\Gamma$ point. The degeneracy of the first and forth branches at $\Gamma$ point is, however, lifted by the non-zero $d_z$, whereas that of the second and third branches remains.  Hence, we infer that our model Hamiltonian, in which we neglect the in-plane DM interaction and the larger-structure of exchange interactions, is not sufficient for the purpose of discussing the scattering intensity of an individual mode. Since the in-plane DM interaction does not resolve the degeneracy at $\Gamma$ point (see Figure S3), the incorporation of the larger-structure of exchange interactions is likely required to settle the intensity issue of the second and third branches. On the other hand, the out-of-plane DM interaction $d_z = 0.18$ determines the wave functions of the first and fourth branches, for which we have no singular behavior around $\Gamma$ point and the larger-structure of exchange interactions is expected to be irrelevant. We note that in Figure S4(b) the fourth branch is as intense as the first branch, which also contradicts the experimental result. This calculated result of the fourth branch with rather strong intensity may suggest that the in-plane DM interaction, whose strength $d_p$ is expected to be the same order as $d_z$, strongly affects the triplet-dimer wave functions. (Note that our omission of the in-plane DM interaction is based on the consideration of excitation energies.) 
	
In summary, there is still an inconsistency between the experimental data and numerical calculations. We believe that this inconsistency is not essential, but is artifact due to the degeneracy at $\Gamma$ point.  Additional terms in the spin Hamiltonian, which resolve the degeneracy of the second and third branches at $\Gamma$ point, could remove the singularity of $S(q,\omega)$ and alter the dynamic structure factors of the second and third branches. Further studies which include such terms, as well as the incorporation of an adequate extrapolation into our perturbation calculation, are apparently necessary to settle this issue. 

\begin{center} {\bf Magnetic domains: clockwise and anticlockwise pinwheel states} \end{center}

We believe that in Rb$_2$Cu$_3$SnF$_{12}$ there are two magnetic domains which give rise two different arrangements of dimers forming clockwise and anticlockwise pinwheel states as shown in Figure S5.  The presence of these two magnetic domains, which are a mirror image of each other, corresponds to two structural domains.  These two different arrangements of dimers generate a distinct scattering pattern.  Figure S6 shows the integrated dynamic structure factor of the clockwise pinwheel state, which after combined with that of the anticlockwise state gives the scattering pattern shown in Figure 2e.

\newpage
\begin{figure}
\vspace{0in}
\centering \includegraphics[width=14cm]{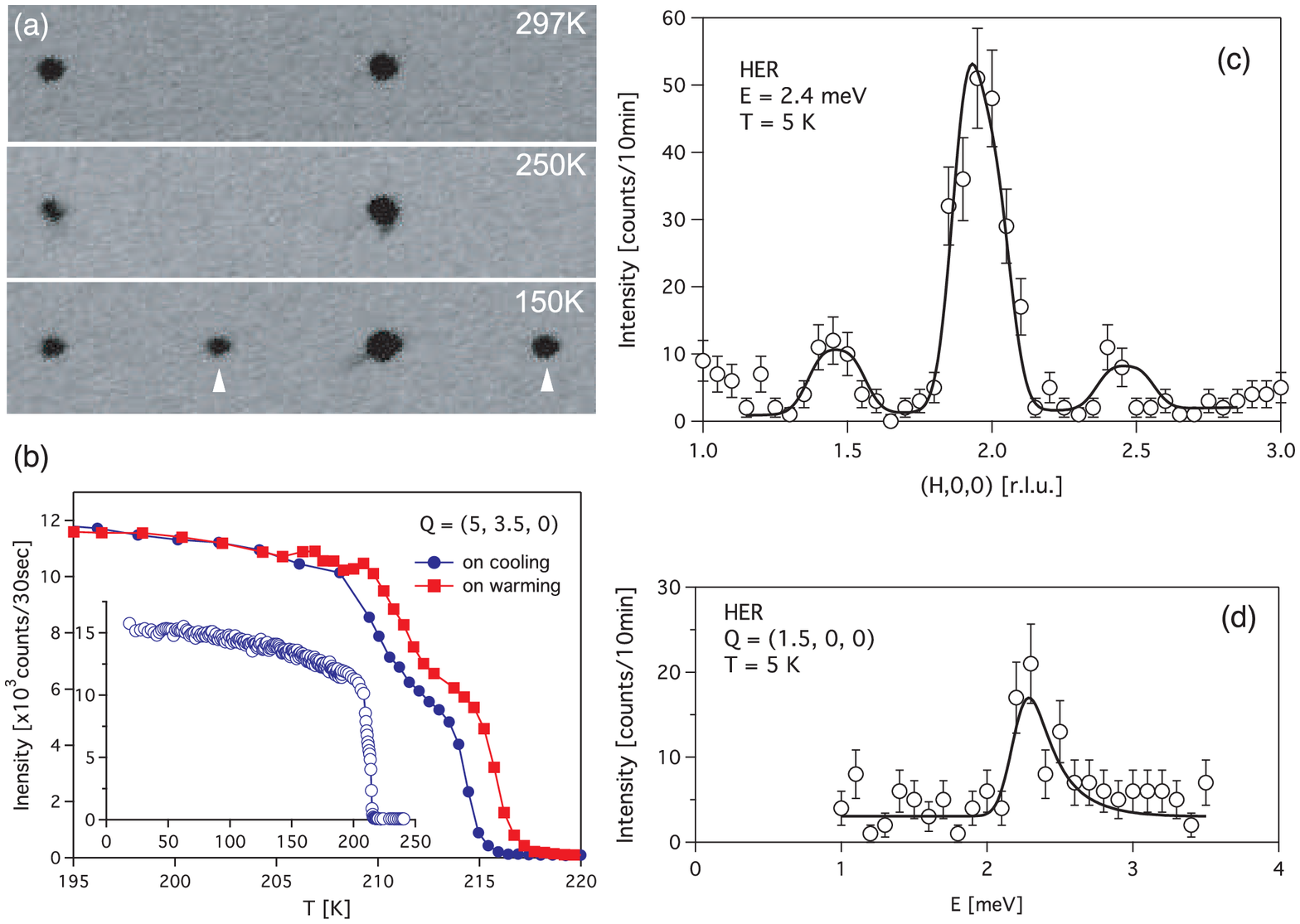}
\begin{flushleft}
Figure S1 (a) X-ray photographs show the fundamental Bragg peaks and the superlattice peaks indicated by the arrows at positions $(\pm1/2,\pm1/2,0)$ away from the fundamental peaks. (b) Intensity of the superlattice peak (5, 3.5, 0) measured with neutron scattering on a single crystal shows the hysteretic behavior at $T_s$. (c) A constant-energy scan at $\hbar\omega = 2.4$ meV reveals the triplet excitations around (1.5, 0, 0), (2, 0, 0) and (2.5, 0, 0).  (d) An energy scan at (1.5, 0, 0) yields $\Delta_1 = 2.21(9)$ meV. Lines are fits to the empirical dispersion relation described in Methods convoluted with the resolution function.
\end{flushleft}
\end{figure}

\begin{figure}
\vspace{0in}
\centering \includegraphics[width=10cm]{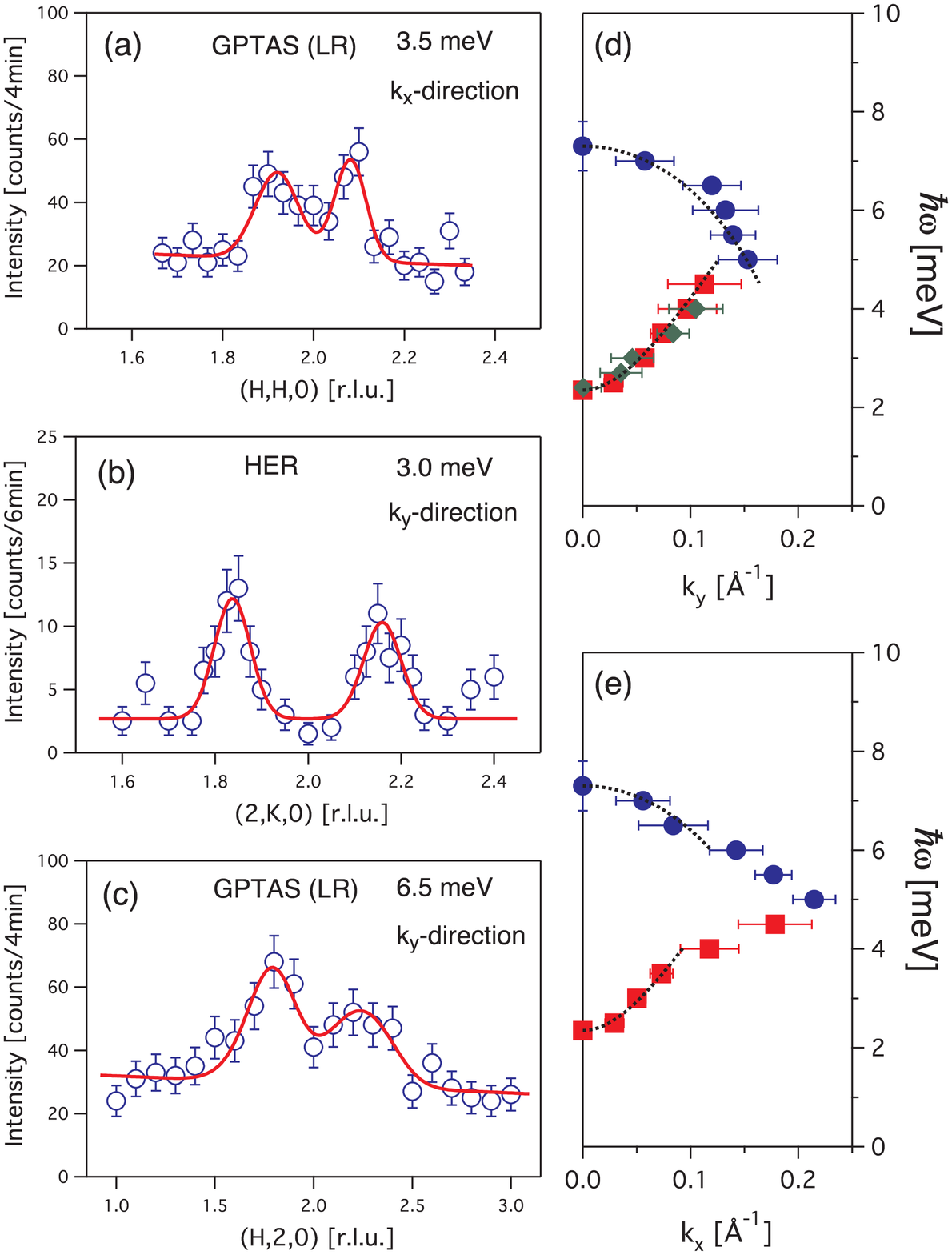}
\begin{flushleft}
Figure S2 (a)-(c) Representative constant-energy scans along $k_x$ and $k_y$. Lines are fits to the empirical dispersion relation described in Methods convoluted with the resolution function. Dotted lines in (d) and (e) show the empirical dispersion relation with a universal $v_\alpha = 35$ meV/\AA.
\end{flushleft}
\end{figure}

\begin{figure}
\vspace{0in}
\centering \includegraphics[width=15cm]{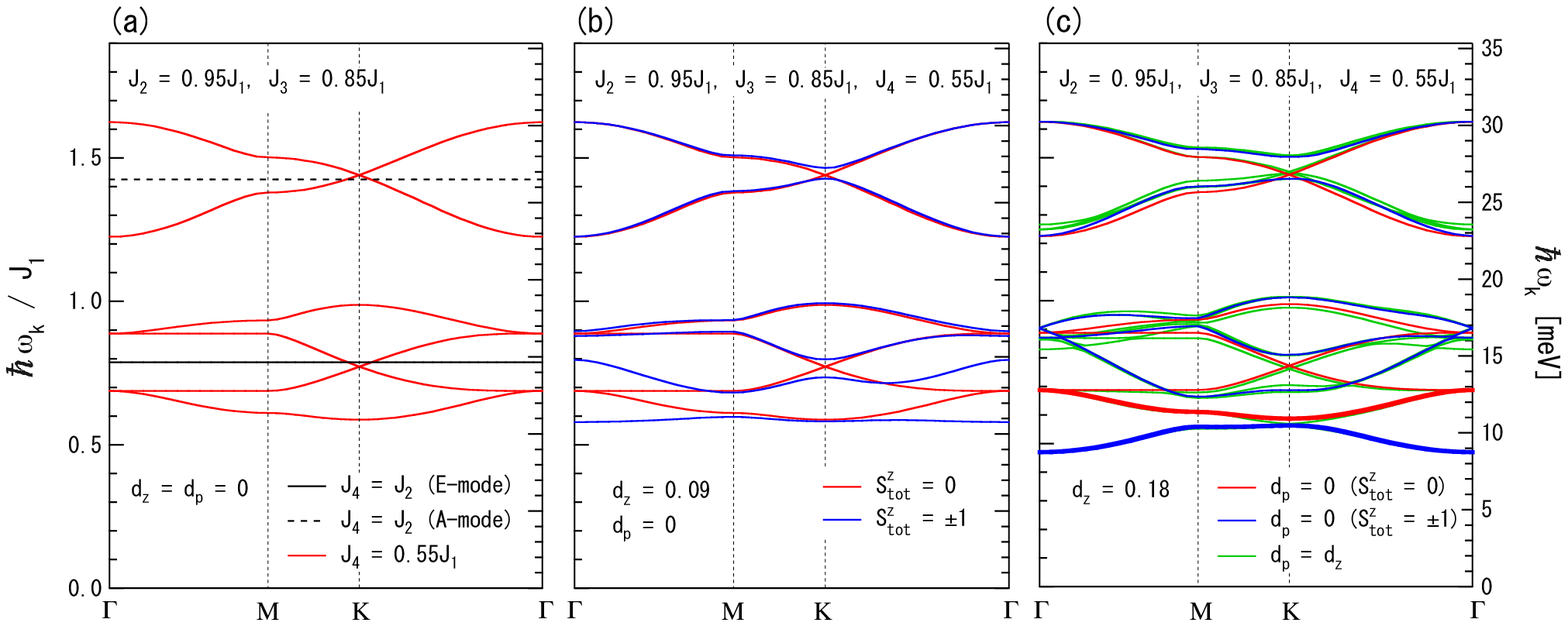}
\begin{flushleft}
Figure S3 First-order results of the triplet-dimer excitation spectra. For the right axis, we use $J_1 = 18.6$ meV.
\end{flushleft}
\end{figure}

\begin{figure}
\vspace{0in}
\centering \includegraphics[width=15cm]{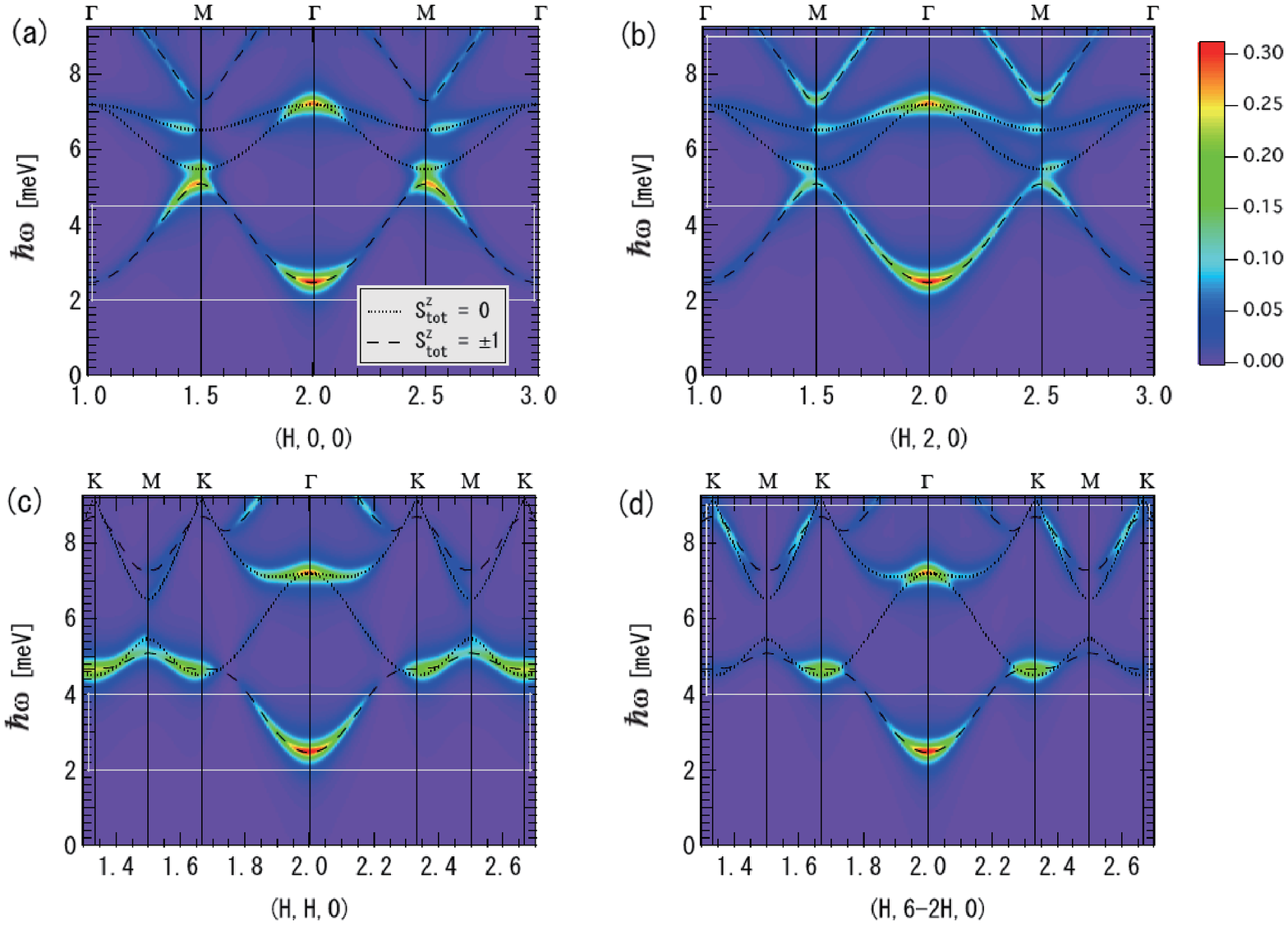}
\begin{flushleft}
Figure S4 Calculated dynamic structure factor $S(q,\omega)$ for the energy range of $\hbar\omega < 9$ meV, where the weight $I_\Lambda$ is estimated by the fifth-order calculation and energy $E_\Lambda$ by  the eighth-order series with the Dlog-Pad{\'e} analysis. White rectangles denote the measuring ranges plotted in Figure 2a and b. 
\end{flushleft}
\end{figure}

\begin{figure}
\vspace{0in}
\centering \includegraphics[width=15cm]{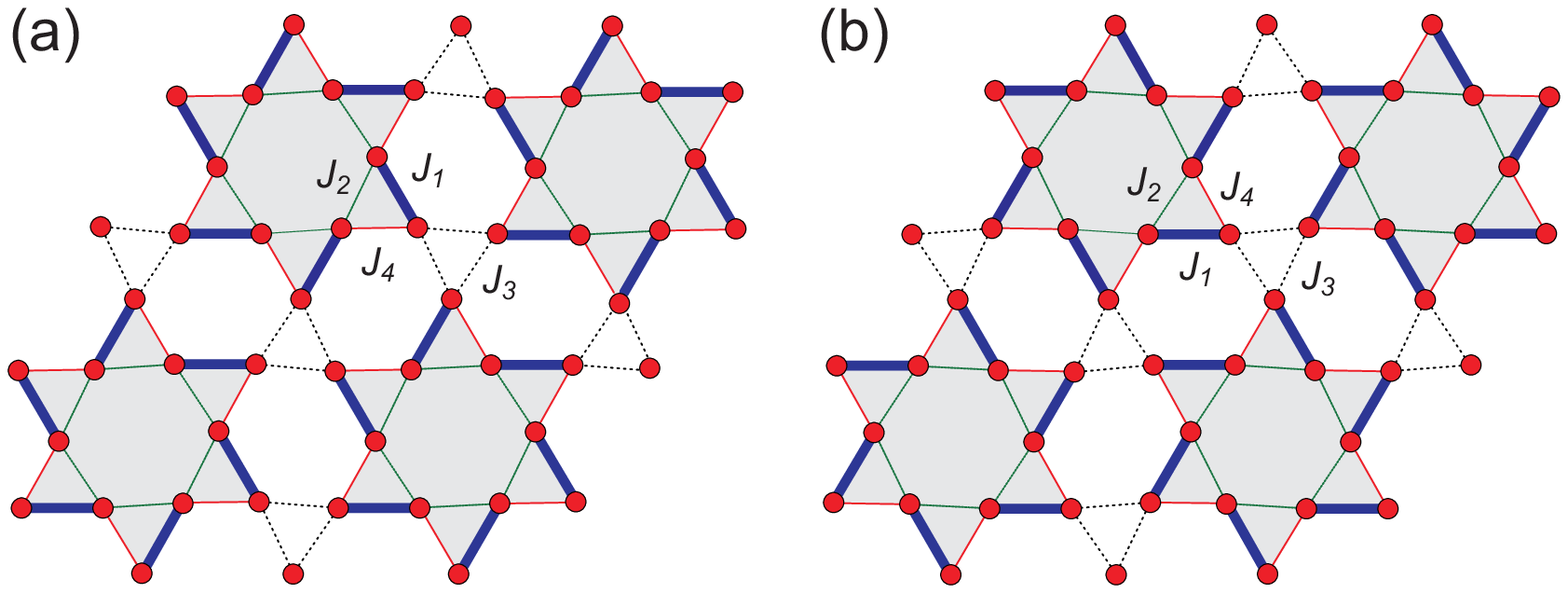}
\begin{flushleft}
Figure S5 Two magnetic domains, (a) a clockwise pinwheel state and (b) an anticlockwise pinwheel state.
\end{flushleft}
\end{figure}
 
\begin{figure}
\vspace{0in}
\centering \includegraphics[width=10cm]{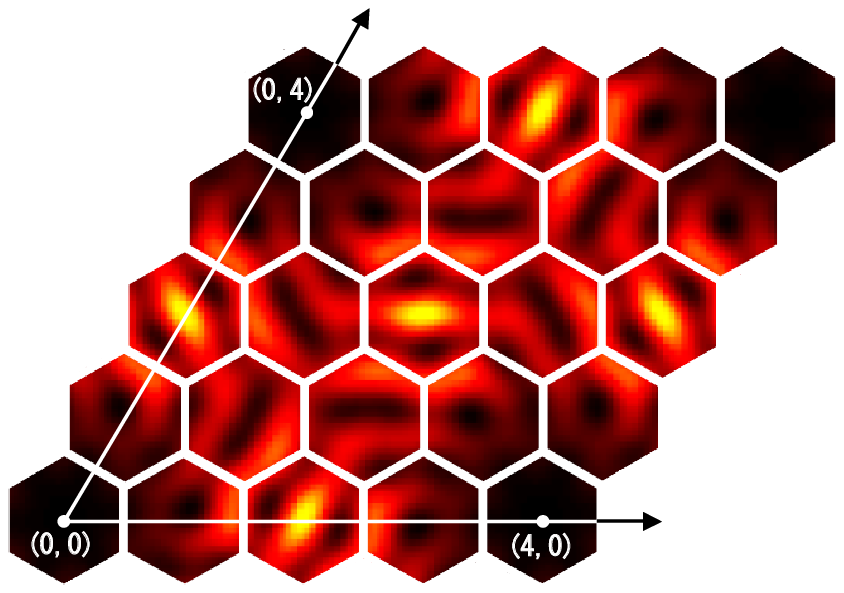}
\begin{flushleft}
Figure S6 The integrated dynamic structure factor is calculated for the clockwise pinwheel state.
\end{flushleft}
\end{figure}

\end{document}